\documentclass[a4paper,11pt]{scrartcl}  % "extarticle" gives more global font size options. use 8pt with Montserrat and 10pt with EBGaramond
\usepackage{import}
\usepackage[left=2cm,top=3cm,bottom=3cm,right=2cm]{geometry}
\usepackage{listings, xcolor}

\definecolor{verylightgray}{rgb}{.97,.97,.97}

\lstdefinelanguage{Solidity}{
	keywords=[1]{anonymous, assembly, assert, balance, break, call, callcode, case, catch, class, constant, continue, constructor, contract, debugger, default, delegatecall, delete, do, else, emit, event, experimental, export, external, false, finally, for, function, gas, if, implements, import, in, indexed, instanceof, interface, internal, is, length, library, log0, log1, log2, log3, log4, memory, modifier, new, payable, pragma, private, protected, public, pure, push, require, return, returns, revert, selfdestruct, send, solidity, storage, struct, suicide, super, switch, then, this, throw, transfer, true, try, typeof, using, value, view, while, with, addmod, ecrecover, keccak256, mulmod, ripemd160, sha256, sha3}, % generic keywords including crypto operations
	keywordstyle=[1]\color{blue}\bfseries,
	keywords=[2]{address, bool, byte, bytes, bytes1, bytes2, bytes3, bytes4, bytes5, bytes6, bytes7, bytes8, bytes9, bytes10, bytes11, bytes12, bytes13, bytes14, bytes15, bytes16, bytes17, bytes18, bytes19, bytes20, bytes21, bytes22, bytes23, bytes24, bytes25, bytes26, bytes27, bytes28, bytes29, bytes30, bytes31, bytes32, enum, int, int8, int16, int24, int32, int40, int48, int56, int64, int72, int80, int88, int96, int104, int112, int120, int128, int136, int144, int152, int160, int168, int176, int184, int192, int200, int208, int216, int224, int232, int240, int248, int256, mapping, string, uint, uint8, uint16, uint24, uint32, uint40, uint48, uint56, uint64, uint72, uint80, uint88, uint96, uint104, uint112, uint120, uint128, uint136, uint144, uint152, uint160, uint168, uint176, uint184, uint192, uint200, uint208, uint216, uint224, uint232, uint240, uint248, uint256, var, void, ether, finney, szabo, wei, days, hours, minutes, seconds, weeks, years},	% types; money and time units
	keywordstyle=[2]\color{teal}\bfseries,
	keywords=[3]{block, blockhash, coinbase, difficulty, gaslimit, number, timestamp, msg, data, gas, sender, sig, value, now, tx, gasprice, origin},	% environment variables
	keywordstyle=[3]\color{violet}\bfseries,
	identifierstyle=\color{black},
	sensitive=true,
	comment=[l]{//},
	morecomment=[s]{/*}{*/},
	commentstyle=\color{gray}\ttfamily,
	stringstyle=\color{red}\ttfamily,
	morestring=[b]',
	morestring=[b]"
}

\usepackage{microtype}

\definecolor{Accent}{rgb}{0.4,0.0,0.0} 
\definecolor{Highlight}{rgb}{0.6,0.0,0.0} 

\definecolor{codegreen}{rgb}{0,0.6,0}
\definecolor{codegray}{rgb}{0.5,0.5,0.5}
\definecolor{codepurple}{rgb}{0.58,0,0.82}
\definecolor{backcolour}{rgb}{0.98,0.98,0.95}
\definecolor{red30}{rgb}{1.0,0.9,0.7}
\definecolor{blue30}{rgb}{0.7,0.9,1.0}

\definecolor{red50}{rgb}{0.7,0.0,0.0}
\definecolor{green50}{rgb}{0.0,0.5,0.0}
\definecolor{yellow50}{rgb}{0.25,0.25,0.0}

\usepackage{orcidlink} % puts a nice little ORCID symbol in the authors' list

\hypersetup{colorlinks,citecolor = Accent, linkcolor = Accent,urlcolor = Accent}

\hypersetup{colorlinks=true}
\hypersetup{linkcolor=red!50!black}
\hypersetup{urlcolor=blue!50!black}
\hypersetup{citecolor=green!50!black}

\hypersetup{pdfsubject={Atomic Settlement}}
\hypersetup{pdfkeywords={Delivery vs Payment, DvP, Settlement, Atomic Swap, Hashed Timelock Contract, HTLC, Smart Contract, Blockchain, ERC 7573}}

\usepackage{amsmath}
\usepackage{wasysym}
\usepackage{textcomp} % some control quote sequences and other symbols
\usepackage{mathtools} % symbols
\usepackage{mathrsfs} %allows the /mathscr command for in-line math
\usepackage{gensymb} % nice \degree 
\usepackage[autostyle]{csquotes} % nice quotes using \enquote{}
\usepackage[english]{babel} % multilingual hyphenation support

\usepackage{array} % allows you to center text in a table with columns of defined width

\usepackage[inline,shortlabels]{enumitem} % in-line enumerated items / resume

\makeatletter
\usepackage{listings}
\lstdefinestyle{mystyle}{
    backgroundcolor=\color{backcolour},   
    commentstyle=\color{codegreen},
    keywordstyle=\color{magenta},
    numberstyle=\tiny\color{codegray},
    stringstyle=\color{codepurple},
    basicstyle=\ttfamily\lst@ifdisplaystyle\scriptsize\fi,
    breakatwhitespace=false,         
    breaklines=true,                 
    captionpos=b,                    
    keepspaces=true,                 
    numbers=left,                    
    numbersep=5pt,                  
    showspaces=false,                
    showstringspaces=false,
    showtabs=false,
    tabsize=2
}
\makeatother

\usepackage{tikz}
\usepackage{pgf-umlsd}
\newenvironment*{messself}[4][1]{
  \stepcounter{seqlevel}
  \stepcounter{callevel} % push
  \stepcounter{callselflevel}

  \path
  (#2)+(\thecallselflevel*0.1-0.1,-\theseqlevel*\unitfactor-0.7*\unitfactor) node (sc\thecallevel) {}
  ({sc\thecallevel}.east)+(0,-0.33*\unitfactor) node (scb\thecallevel) {};

  \draw[->,>=triangle 60] ({sc\thecallevel}.east) -- ++(0.8,0)
  node[near start, above right] {#3} -- ++(0,-0.33*\unitfactor)
  -- (scb\thecallevel) node[near end, below right] {#4};
  \def\l\thecallevel{#1}
  \def\f\thecallevel{#2}
  \def\t\thecallevel{#2}
  \def\returnvalue{#4}
  \tikzstyle{threadstyle}+=[instcolor#2]
}{
  \addtocounter{callevel}{-1} % pop
  \addtocounter{callselflevel}{-1}
}
\newenvironment{messcalldashed}[4][1]{
  \stepcounter{seqlevel}
  \stepcounter{callevel} % push
  \path
  (#2)+(0,-\theseqlevel*\unitfactor-0.7*\unitfactor) node (cf\thecallevel) {}
  (#4.\threadbias)+(0,-\theseqlevel*\unitfactor-0.7*\unitfactor) node (ct\thecallevel) {};
  
  \draw[dashed,->,>=angle 60] ({cf\thecallevel}) -- (ct\thecallevel)
  node[midway, above] {#3};
  \def\l\thecallevel{#1}
  \def\f\thecallevel{#2}
  \def\t\thecallevel{#4}
  \tikzstyle{threadstyle}+=[instcolor#2]
}
{
  \addtocounter{seqlevel}{\l\thecallevel}
  \path
  (\f\thecallevel)+(0,-\theseqlevel*\unitfactor-0.7*\unitfactor) node (rf\thecallevel) {}
  (\t\thecallevel.\threadbias)+(0,-\theseqlevel*\unitfactor-0.3*\unitfactor) node (rt\thecallevel) {};
  \drawthread{ct\thecallevel}{rt\thecallevel}
  \addtocounter{callevel}{-1} % pop
}

\usepackage{rotating}

\newcommand{\articletitle}{A Stateless and Secure Delivery versus Payment across two Blockchains}

\title{Secure Cross-Chain Delivery-vs-Payment\\ using Stateless Decryption Oracles}

\title{Stateless and Secure\\ Delivery versus Payment across Blockchains}

\hypersetup{pdftitle={\articletitle}}
\hypersetup{pdfauthor={Christian P. Fries, Peter Kohl-Landgraf}}

\author{%
    Christian Fries\orcidlink{0000-0003-4767-2034}{\textsuperscript{1,2,*}}
    \and
    Peter Kohl-Landgraf{\textsuperscript{1,*}}
}

\date{November 9th, 2023}

\begin{document}

\thispagestyle{empty}

\maketitle

\vspace{-1.25cm}
\begin{center}
{\small (version 3.6 [v7])}
\end{center}

\begin{center}
    \begin{small}
%        \textcolor{Accent
{\textsuperscript{1}}DZ BANK AG, Deutsche Zentral-Genossenschaftsbank, Frankfurt a. M., Germany\\ 
%        \textcolor{Accent}
{\textsuperscript{2}}Department of Mathematics, Ludwig Maximilians University, Munich, Germany\\
%        \textcolor{Accent}
{\textsuperscript{*}}Correspondence: %\textcolor{Accent}
{\href{mailto:email@christian-fries.de}{email@christian-fries.de}, \href{mailto:pekola@mailbox.org}{pekola@mailbox.org}} \\
    \end{small}
\end{center}

%%%%%%%%%%%%%%%%%%%%%%%%%%%%%%%%%%%%%%%%%%%%%%%%%%%%%%%%%%%%%%%%%%%%%
% Abstract
%%%%%%%%%%%%%%%%%%%%%%%%%%%%%%%%%%%%%%%%%%%%%%%%%%%%%%%%%%%%%%%%%%%%%

\section*{Abstract}

\noindent

We propose a \textbf{secure, stateless and composable transaction scheme} to establish delivery-versus-payment (DvP) across two (or more) blockchains \textbf{without relying on time-locks, centralized escrow, or stateful intermediaries}. The method minimizes coordination overhead and removes race conditions via a stateless decryption oracle that conditionally releases cryptographic keys.

Specifically, the scheme requires:
\begin{enumerate}
    \item a decryption oracle service—either centralized or using threshold decryption—that decrypts transaction-specific encrypted messages, and
    
    \item a ``payment contract'' on the payment chain that executes conditional payments via
    \begin{lstlisting}[numbers=none]
    transferAndDecrypt(uint256 id, address from, address to,
                        string keyEncryptedSuccess, string keyEncryptedFail)    \end{lstlisting}
    and emits the appropriate key, depending on transaction outcome.
\end{enumerate}
The decrypted key then deterministically enables follow-up transactions - such as asset delivery or cancellation - on a separate blockchain.

The protocol is lightweight and compatible with existing blockchain infrastructure (e.g., Ethereum), and avoids timeouts or pre-defined orderings. Our approach improves atomic cross-chain settlement and can serve as a blueprint for decentralized inter-chain financial markets.

The protocol allows for multi-party DvP across multiple chains. A multi-party delivery versus payment is a valuable trade feature as it allows to bound multiple trade into a single atomic unit, effectively reducing liquidity requirements.

\subsubsection*{Disclaimer}

The views expressed in this work are the personal views of the presenters and do not necessarily reflect the views or policies of current or previous employers.

\clearpage

\setcounter{tocdepth}{2}

\begin{footnotesize}
\tableofcontents    
\end{footnotesize}

\clearpage

%\begin{multicols*}{2} % if you want two columns
%%%%%%%%%%%%%%%%%%%%%%%%%%%%%%%%%%%%%%%%%%%%%%%%%%%%%%%%%%%%
\section{Introduction}
%%%%%%%%%%%%%%%%%%%%%%%%%%%%%%%%%%%%%%%%%%%%%%%%%%%%%%%%%%%%

Delivery-versus-Payment (DvP), the atomic settlement of an asset in exchange for payment, is a fundamental requirement in both traditional financial markets and blockchain-based systems. For tokenized assets, stablecoins, or CBDCs on separate ledgers, ensuring secure DvP without trusted intermediaries poses a persistent challenge.

Conventional solutions, such as Central Counterparty Clearinghouses (CCPs) or Central Securities Depositories (CSDs), mitigate settlement risk but rely on centralized infrastructure, require collateral, and introduce latency and cost \cite{bis1992dvp}. In the context of distributed ledger technology (DLT) \cite{Bechtel2022}, several proposals, including Hashed Timelock Contracts (HTLCs) and API-based DvP mechanisms, offer decentralized alternatives, but suffer from limitations such as race conditions, reliance on timeouts and high on-chain complexity \cite{ECB2018, BancaItalia}. The proposed solutions are based on a stateful, often API-based interaction mechanism with centralized trusted entities \cite{BancaItalia,Zakhary_2020}, which bridges the communication between a so-called Asset Chain and a corresponding payment system, or require time-based constraints. These time-based constraints could be time-locks, which are required at least on one of the two chains, \cite{lee2021atomic, LightSwapAtomcSwap2023}, or verifiable timed signatures (VTS) \cite{thyagarajan2022universal}. Time-based constraints can be problematic, as they introduce short-term options, which will ultimately be part of the transaction cost, and still exhibit race conditions.

In this paper, we address the core technical challenge of cross-chain DvP: avoiding front-running and race conditions in function calls across two (or more) ledgers, while ensuring that neither asset nor payment is lost or duplicated. Our contribution is a protocol that eliminates the need for trusted intermediaries, time constraints (time-locks or VTS), or persistent off-chain state.

We propose a stateless decryption oracle-based approach, in which encrypted keys are embedded into smart contracts and decrypted conditionally based on transaction outcome. This ensures atomicity and fairness while preserving decentralization. Moreover, the scheme is compatible with threshold decryption, enabling decentralized oracle architectures and minimizing trust assumptions.

\medskip

For settlement involving central bank digital currency or commercial bank money tokens \cite{braine2024paymentsusecasesdesign}, it may be reasonable to integrate the decryption oracle into the payment system. However, this is not a requirement.

\medskip

The smart contracts proposed here are available as ERC 7573 \cite{erc7573} in the Ethereum Improvement Process, \cite{ethereum_ercs}.

\subsection{Contribution}

Unlike HTLC-based protocols, our approach eliminates time-locks entirely, removing the need for enforced expiration conditions and mitigating associated race conditions. Our scheme also avoids costly on-chain cryptographic operations, relying instead on off-chain verification via a stateless oracle. With respect to on-chain verification, our approach does not require complex signing or encryption algorithms to run on-chain.
As a by-product, we elaborate that the central problem in cross-blockchain DvP is that function arguments are observable before the function execution is final. We show that the use of decryption oracles solves this general problem; see Figure~\ref{fig:dvp:decryptionoracle}.

\subsection{Comparison of Delivery-versus-Payment Mechanisms}

The following table gives a brief comparison of the present method and the most common DvP methods.
\begin{table}[h]
    \renewcommand{\arraystretch}{1.5}
    \centering
    \tiny
        \begin{tabular}{|p{0.17\textwidth}|p{0.17\textwidth}|p{0.17\textwidth}|p{0.17\textwidth}|p{0.17\textwidth}|}
        \hline
        \raggedright\textbf{Criterion}\arraybackslash & \textbf{Proposed Method} & \textbf{HTLCs} & \textbf{Centralized DvP} & \textbf{API-Based DvP} \\ 
        \hline
        \raggedright\textbf{Intermediary Required?} & No & No & Yes & Yes \\ 
        \hline
        \raggedright\textbf{External Storage Required?} & No & Yes (storage for time-lock) & Yes & Yes \\ 
        \hline
        \raggedright\textbf{Timeout Required?} & No & Yes & No & No \\ 
        \hline
        \textbf{Coupling of Payment \& Delivery} & Logically ordered via cryptographic key release & Coupled via hash preimage and timeing & Coupled via intermediary & Coupled via API function \\ 
        \hline
        \raggedright\textbf{Flexibility in Execution} & High (stateless, no fixed time) & Medium (timeout can be problematic) & Low (dependent on central system) & Medium (dependent on API availability) \\ 
        \hline
        \raggedright\textbf{On-Chain Costs} & Low (no time-locks, stateless) & Medium (gas fees apply to time-locks) & No direct on-chain costs, but high service fees & Medium (depends on API fees) \\        
        \hline
        \raggedright\textbf{Security Risks} & Low (no central entity) & Medium (reorgs, time-lock attacks possible) & High (centralization introduces risks) & Medium (depends on API integrity) \\ 
        \hline
    \end{tabular}
    \caption{Comparison of DvP Methods}
    \label{tab:dvp_comparison}
\end{table}

\bigskip

The rest of this paper is structured as follows: Section~\ref{sec:dvp:problemdescription} formalizes the problem; Section~\ref{sec:dvp:contractinterfaces} introduces the smart contract interfaces; Section~\ref{sec:dvp:workflow} details the full DvP workflow; Section~\ref{sec:dvp:decryptionoraclekeyformat} discusses the decryption oracle design, including threshold schemes; Section~\ref{sec:dvp:remarks} provides security and implementation considerations; Section~\ref{sec:dvp:multiparty} describes the extension to a multi-party DvP; and Section~\ref{sec:dvp:conclusion} concludes.

%%%%%%%%%%%%%%%%%%%%%%%%%%%%%%%%%%%%%%%%%%%%%%%%%%%%%%%%%%%%
\subsection{Acknowledgments}
%%%%%%%%%%%%%%%%%%%%%%%%%%%%%%%%%%%%%%%%%%%%%%%%%%%%%%%%%%%%

We like to thank Giuseppe Galano, Julius Lauterbach and Stephan M\"{o}gelin for their valuable feedback.

\pagebreak[5]
%%%%%%%%%%%%%%%%%%%%%%%%%%%%%%%%%%%%%%%%%%%%%%%%%%%%%%%%%%%%
\section{Problem Description}
\label{sec:dvp:problemdescription}
%%%%%%%%%%%%%%%%%%%%%%%%%%%%%%%%%%%%%%%%%%%%%%%%%%%%%%%%%%%%

\paragraph{Problem Statement.} \textit{
Given two blockchain networks (e.g., AssetChain and PaymentChain), and a buyer/seller pair wishing to perform a delivery-versus-payment (DvP) exchange without a trusted intermediary, how can one ensure atomicity and fairness of settlement, while avoiding race conditions, observable arguments, reliance on timeouts, and external state?
}

\bigskip

The core issue in cross-chain DvP is a \textit{race condition} caused by observable function arguments prior to execution (\textit{function argument leakage before function finality}). This problem is specific to the distributed nature of code execution in blockchains and likely not present in classical trusted environments.

To illustrate this, consider two contracts, \textit{AssetContract} and \textit{PaymentContract}, operating on separate blockchains: \textit{AssetChain} and \textit{PaymentChain}. Although these names suggest a specific use case, the two chains could manage tokens representing different types of assets. We aim to perform a transfer on the \textit{AssetChain} if and only if a transfer on the \textit{PaymentChain} was successful.

In addition, consider two participants, the \textit{buyer} and the \textit{seller}. The buyer is receiving the asset and delivering the payment. The seller is delivering the asset and receiving the payment.

Contracts can store values and calculate hashes, thus, compare hashes of given arguments with previously stored values. In addition, contracts can verify the caller of a function.

Assume that there is a sequence of function calls that brings the two contracts into a double-locking state that is as follows:\footnote{We show how to create this situation in Section~\ref{sec:dvp:doublelocking}.}
\begin{itemize}
    \item On the asset contract, the asset has been transferred from the seller to a ``lock'', and
    \begin{itemize}
        \item if the buyer calls the function \texttt{AssetContract::transferWithKey(K)} with $K = B$, the asset will be transferred to the buyer, or

        \item if the seller calls the function \texttt{AssetContract::transferWithKey(K)} with $K = S$, the asset will be transferred to the seller.
    \end{itemize}

    \item On the payment contract, the payment has been transferred from the buyer to a ``lock'', and
    \begin{itemize}
        \item if the seller calls the function \texttt{PaymentContract::transferWithKey(K)} with $K = B$, the payment will be transferred to the seller, or

        \item if the buyer calls the function \texttt{AssetContract::transferWithKey(K)} with $K = S$, the payment will be transferred to the buyer.
    \end{itemize}

    \item The seller knows $B$, but does not know $S$.

    \item The buyer knows $S$, but does not know $B$.
\end{itemize}

It appears as if this situation solves the secure delivery-versus-payment, because
\begin{itemize}
    \item if the seller collects the payment, the buyer observes $B$ an can collect the asset (completion of the transaction), or

    \item if the buyer re-collects the payment, the seller observes $S$ and can re-collect the asset (cancellation of the transaction).
\end{itemize}

However, the above situation (and also its creation) suffers from a \emph{race-condition}:

It must be ensured that the function \texttt{PaymentContract::transferWithKey(K)} can be called only once. So once it is called by one party, it is blocked by the other party. Now, assume that both parties call this function at the same time. One call goes through; the other has to be blocked. If the function performs the blocking interally, the argument $K$ can be observed on the blocked call. Hence, it may be possible that both keys $S$ and $B$ are observed, which clearly compromises the scheme.

For example, the seller likes to complete the transaction and calls \texttt{PaymentContract::transferWithKey(B)}, but right before the buyer called \texttt{PaymentContract::transferWithKey(S)}, which transferred the payment back to the buyer. The seller's call is unsuccessful, but the argument $B$ has been observed, allowing the buyer to fetch the asset (without a payment), given that he is fast enough on the asset chain; see Figure~\ref{fig:dvp:racecondition}. 

\begin{figure}[hbtp]
    \centering
    \scalebox{0.65}{
  \begin{sequencediagram}[instwidth=10cm, instsep=4cm]
    \tikzstyle{inststyle}+=[bottom color=white, top color=white, rounded corners=3mm]
    \newthread[blue30]{s}{Seller of Asset}{}
    \tikzstyle{inststyle}+=[bottom color=white, top color=white, rounded corners=0mm]
    \newinst[2]{p}{Payment Contract}{}
    \tikzstyle{inststyle}+=[bottom color=white, top color=white, rounded corners=3mm]
    \newthread[red30]{b}{Buyer of Asset}{} 
    \tikzstyle{inststyle}+=[bottom color=white, top color=white, rounded corners=0mm]
    \newinst[2]{c}{Asset Contract}{}

    \draw[red, very thick] ([yshift=-1.0cm]p) circle [radius=0.5cm];
    \node[red] at ([xshift=-0.6cm,yshift=-1.2cm]p) {\lightning};
    
    \begin{call}{s}{\textbf{\color{black}transferWithKey(B)}}{p}{\shortstack{\textit{\color{red}Rejected} \\ (but key $B$ was exposed)}}
    \postlevel
    \postlevel
    \postlevel
    \postlevel
    \end{call}

    \prelevel   
    \prelevel   
    \prelevel
    \prelevel
    \prelevel
    \prelevel

    \begin{call}{b}{\textbf{\color{black}transferWithKey(S)}}{p}{\shortstack{\textit{\color{green50}Successful} \\(cancel money transfer)}}

    \begin{call}{b}{\textbf{\color{black}transferWithKey(B)}}{c}{\shortstack{\textit{\color{green50}Successful} \\(collect asset)}}
    \postlevel
    \end{call}

    \end{call}
    \postlevel

  \end{sequencediagram}
    }
    \caption{Race condition in cross-chain DvP: After a near simultaneous call to the payment contract (red circle), the key $B$ can be observed in the mempool or failed function call and then misused on the other chain. The buyer is canceling its payment (with $S$) \emph{and} collecting the asset (with $B$).
    }
    \label{fig:dvp:racecondition}
\end{figure}

\subsection{Limitations of Time Locks}

The above problem can be partially solved by \textit{time-locks}, which, however, have similar vulnerabilities.

\sloppypar With a time-lock, the seller can collect the payment with a call to \texttt{PaymentContract::transferWithKey(B)} if and only if he performs this call within a specific time. After some pre-agreed time $T_{1}$ the payment will be transferred back to the buyer if not collected. The buyer has no option to collect/cancel the payment before (that is, there is no cancellation key $S$).

Likewise, there is a time-lock on the asset chain that transfers the asset back to the seller after time $T_{2}$, with $T_{2} > T_{1}$.

\sloppypar This setup still exhibits the same race condition: if the seller makes his call to \texttt{PaymentContract::transferWithKey(B)} slightly too late, he fails to collect the payment, but exposes $B$ (allowing the buyer to collect the asset).

In addition, if the seller successfully collects the payment, the buyer may still fail to collect the asset if technical issues prevent him from completing his call before $T_{2}$. Hence, time-locks are susceptible to denial-of-service attacks, preventing one party from completing its function call within a given time-frame.

\subsection{Decryption Oracle to avoid Key Exposure}

A secure solution to the race condition problem is the introduction of \emph{encrypted arguments} and a secure decryption oracle.

Assume that there is an offchain oracle that performs a decryption service exclusively for specific contract.

Encryption of a document $K$ can be performed with a public key, but decryption of an encrypted document $E(K)$ can only be requested by the specific contract.

In that case, we may implement a \texttt{function} in a way that consumes an encrypted argument \texttt{argumentEncrypted}, performs the necessary synchronization steps to prevent a race condition, i.e., checks if an encryption has already been performed, and then and only then, request decryption of the given argument. The processing of the decrypted argument is then continued in a callback. Alternatively, synchronization could also take place in the decryption oracle, by preventing exposing a key that belongs to a transaction for which a key has already been exposed.

\begin{figure}[hbtp]
    \centering
    \scalebox{0.55}{
  \begin{sequencediagram}
    \tikzstyle{inststyle}+=[bottom color=white, top color=white, rounded corners=3mm]
    \newthread[blue30]{u}{User}{}
    \tikzstyle{inststyle}+=[bottom color=white, top color=white, rounded corners=0mm]
    \newinst[3]{c}{\shortstack{Contract}}{}
    \newinst[5]{o}{\shortstack{Decryption Oracle}}{}

    \begin{sdblock}{function}{(encrypted arguments)}

    \begin{call}{u}{\textbf{function}(\textbf{\color{blue}argEncrypted})}{c}{\textit{}}

    \begin{call}{c}{preconditionsCheck()}{c}{status}
    \end{call}
    
    \begin{sdblock}{Alt: if status valid}{}

        \begin{messcalldashed}{c}{\textit{DecryptionRequested(id, \textbf{\color{blue}argEncrypted)}}}{o}
            \postlevel
            \begin{messself}{o}{\shortstack[l]{\textbf{\color{red50}decrypt} \textbf{\color{blue}argEncrypted}\\ with private key}}{\textbf{\color{green50}arg}}
            \end{messself}
            \begin{messcall}{o}{\textbf{continue}(id,\textbf{\color{green50}arg})}{c}
            \end{messcall}
            \prelevel
        \end{messcalldashed}

        \begin{messself}{c}{perform operation with \textbf{\color{green50}arg}}{}
        \end{messself}

    \begin{messcalldashed}{c}{Success}{u}
    \end{messcalldashed}

    \prelevel
    \end{sdblock}
    
    \begin{sdblock}{Alt: else}{}

        \prelevel
        \begin{messself}{c}{}{}
        \end{messself}
        
        % Small hack. Emit key to both
        \begin{messcalldashed}{c}{Failure}{u}
        \end{messcalldashed}
    
    \prelevel
    \prelevel
    
    \end{sdblock}

    \prelevel

    \end{call}

    \end{sdblock}

  \end{sequencediagram}
    }
    \caption{Example for a function with encrypted arguments interacting with a decryption oracle.
    }
    \label{fig:dvp:decryptionoracle}
\end{figure}

\subsection{Avoiding Oracle Centralization - Decentralization and Disintermediation}

It appears as if the requirement of a central trusted decryption oracle annihilates the advantages of a blockchain or distributed ledger with respect to decentralization and disintermediation.
However, there could be many equivalent decryption oracle services attached to the network, and counterparties could agree on a decryption oracle per transaction basis. Therefore, the decryption oracle does not represent a monopolistic central role. It is an interchangeable service.

\subsection{Threat Model}

We assume that the two interacting blockchains (AssetChain and PaymentChain) are honest but asynchronous. That is, transactions may be delayed, reordered, or temporarily withheld from inclusion, but are not corrupted.

Participants (buyer and seller) may act adversarially to gain the asset or payment without completing the full protocol. In particular, we assume that:

\begin{itemize}
    \item An adversary can monitor the public mempool or observe in-flight transactions (i.e., arguments submitted to a contract are publicly visible before final inclusion).
    \item Either party may front-run or abort transactions at strategic moments (e.g., after seeing a key B or S on one chain).
    \item No off-chain coordination channel exists that is guaranteed to be reliable or secure.
    \item Smart contracts behave as programmed and enforce access controls and hash comparisons reliably.
    \item The decryption oracle (or its threshold-decryption implementation) only releases keys under verifiable conditions specified in the encrypted payload.
\end{itemize}

Our goal is to prevent either party from gaining both the asset and the payment (break of atomicity), or from unfairly denying the counterparty their rightful claim or refund. We focus especially on mitigating race conditions due to observable keys and asynchrony in cross-chain calls.

\subsection{Why Statelessness Matters}
\label{sec:dvp:introduction:stateless}

A key design goal of our approach is to avoid any reliance on persistent state held by off-chain services. Stateful intermediaries, such as notary contracts, watchtowers, oracles with memory, or external coordination servers, can reintroduce points of trust, synchronization bottlenecks, and denial-of-service risks.

In particular, any solution that requires:
\begin{itemize}
    \item tracking per-transaction metadata off-chain,
    \item storing ephemeral keys for later release,
    \item or maintaining mutable state across multiple parties or chains,
\end{itemize}
violates the principle of decentralized trust minimization.

Instead, we employ a \emph{stateless decryption oracle}, which makes decisions to be based solely on the content of the encrypted message and the verified contract call. This eliminates external dependencies and enables verifiable, deterministic key release, without the need for external memory or coordination.

This statelessness also improves scalability and modularity: any party can deploy a decryption oracle, and counterparties can mutually agree on a trusted instance or use threshold decryption to distribute trust.

We formalize and implement this approach in Section~\ref{sec:dvp:decryptionoraclekeyformat}.

\clearpage
%%%%%%%%%%%%%%%%%%%%%%%%%%%%%%%%%%%%%%%%%%%%%%%%%%%%%%%%%%%%
\section{Contract Interfaces}
\label{sec:dvp:contractinterfaces}
%%%%%%%%%%%%%%%%%%%%%%%%%%%%%%%%%%%%%%%%%%%%%%%%%%%%%%%%%%%%

Consider the setup of having two chains managing two different types of tokens. An example is a chain managing tokenized assets (the \textit{Asset Chain}) and a chain that allows to trigger and verify payments (\textit{Payment Chain}).

We define two interfaces, \texttt{ILockingContract} and \texttt{IDecryptionContract}, to decouple asset delivery from payment confirmation. This separation allows independent verification of delivery and payment, coordinated only via cryptographic key release.
\begin{itemize}
    \item \texttt{ILockingContract}: a smart contract implementing this interface is able to lock the transfer of a token. The transfer can be completed by presentation of a success key ($B$), or reverted by presentation of a failure key ($S$). Presentation of $B$ transfers the token to the buyer, presentation of $S$ re-transfers the token back to the seller.

    \item \texttt{IDecryptionContract}: a smart contract implementing this interface offers a transfer method that performs a conditional decryption of one of two encrypted keys ($E(B)$, $E(S)$), conditional on success or failure.
\end{itemize}
For decryption, we propose a decryption oracle that offers a stateless service for verification and decryption of encrypted keys and, for convenience, can also perform the generation of the keys.

While the \texttt{IDecryptionContract}'s conditional transfer is prepared with encryptions $E(B)$, $E(S)$ of the keys $B$, $S$, the \texttt{ILockingContract}'s conditional transfer is prepared with hashes utilizing a hashing $H(B)$, $H(S)$ of the keys $B$, $S$.
This may be useful as hashing is a comparably cheap operation, while on-chain encryption may be costly.

To verify the consistency of the conditional transfers of the two contracts, the decryption oracle offers a method that allows one to obtain $H(K)$ from a given $E(K)$ without exposing $K$.

\smallskip

If on-chain encryption is cheap, the hashes may be replaced with the encrypted keys, $H(K) = E(K)$, which slightly simplifies the protocol. We will describe the general case.

\medskip

Neither contract maintains long-term state about transactions; correctness depends on hash/key matching. The trust model is reduced to ensuring keys are unique per transaction and correctly verified, which is enforced via the decryption oracle.

\subsection{Notation}

The method proposed here relies on a service that will decrypt an encrypted document, where the encrypted document is observed in a message emitted by the smart contract implementing \texttt{IDecryptionContract} on the payment chain. In the following, we call these documents \textit{key}, because they are used to unlock transactions (on the contact implementing \texttt{ILockingContract}).

We call the receiver of the tokens handled by \texttt{ILockingContract} the \textit{buyer}, and the payer of the tokens handled by the \texttt{ILockingContract} the \textit{seller}. For tokens handled by the \texttt{IDecryptionContract}, the flow is reversed, the aforementioned \textit{buyer} is the payer of the \texttt{IDecryptionContract}'s tokens, and the aforementioned \textit{seller} is the receiver of the \texttt{IDecryptionContract}'s tokens. This fits to the interpretation that the tokens on the \texttt{IDecryptionContract} are a payment for the transfer of the tokens on the \texttt{ILockingContract}.

There is a key for the buyer and a key for the seller. In Figure~\ref{fig:deliveryvspayment:sequence:full} and throughout, these keys are denoted by $B$ (buyer key, successful payment) and $S$ (seller key, failed payment), respectively. The encrypted keys are denoted by $E(B)$ and $E(S)$ and hashes of the keys are denoted by $H(B)$ and $H(S)$.

\subsection{\texttt{ILockingContract}}
\label{sec:dvp:ILockingContract}

The interface \texttt{ILockingContract} is given by the following methods:

\begin{lstlisting}[language=Solidity]
    inceptTransfer(uint256 id, int amount, address from, string memory keyHashedSeller, string memory keyEncryptedSeller);

    confirmTransfer(uint256 id, int amount, address to, string memory keyHashedBuyer, string memory keyEncryptedBuyer);

    cancelTransfer(uint256 id, int amount, address from, string memory keyHashedSeller, string memory keyEncryptedSeller);

    transferWithKey(uint256 id, string key);
\end{lstlisting}

A contract implementing this interface provides a transfer of tokens (usually presenting the delivery of an asset), where the tokens are temporarily locked. The completion or reversal of the transfer is then conditional on the presentation of one of the two keys.
\begin{itemize}
    \item \lstinline{inceptTransfer}: Called by the buyer of the token whose address is implicit (\lstinline{to = msg.sender}). Sets the hash of a key that will revert the token to the seller (\lstinline{from}).

    \item \lstinline{confirmTransfer}: Called by the seller of the token whose address is implicit (\lstinline{from = msg.sender}). Verifies that the seller's (\lstinline{from}) and buyer's (\lstinline{to}) addresses match the corresponding call (with the same (\lstinline{id})) to \lstinline{inceptTransfer}. Sets the hash of a key that will trigger transfer of the token to the buyer (\lstinline{to}).

    \item \lstinline{transferWithKey}: Called by the buyer or seller of the token with a \lstinline{key} whose hash matches the \texttt{keyHashedBuyer} or \texttt{keyHashedSeller} respectively (which ever key is released by the \texttt{IDecryptionContract}).
\end{itemize}

\subsection{\texttt{IDecryptionContract}}
\label{sec:dvp:IDecryptionContract}

The interface \texttt{IDecryptionContract} is given by the following methods:

\begin{lstlisting}
    inceptTransfer(uint256 id, int amount, address from, string memory keyEncryptedSuccess, string keyEncryptedFailure);

    transferAndDecrypt(uint256 id, int amount, address to, string memory keyEncryptedSuccess, string keyEncryptedFailure);

    cancelAndDecrypt(uint256 id, address from, string memory keyEncryptedSuccess, string memory keyEncryptedFailure);

    releaseKey(uint256 id, string memory key) external;
\end{lstlisting}

A contract implementing this interface provides a transfer of tokens (usually representing a payment), where a successful or failed transfer releases one of two keys, respectively.
\begin{itemize}
    \item \lstinline{inceptTransfer}: Called by the receiver of the (payment) token whose address is implicit (\lstinline{to = msg.sender}). Sets the encrypted keys that will be decrypted upon the success or failure of the transfer from the payer (\lstinline{from}).

    \item \lstinline{transferAndDecrypt}: Called by the payer of the (payment) token whose address is implicit (\lstinline{from = msg.sender}). 
    Verifies that the payer's (\lstinline{from}) address, receiver's (\lstinline{to}) address, and the encrypted keys match the corresponding call (with the same \lstinline{id}) to \lstinline{inceptTransfer}.
    Tries to perform a transfer of the token.
    
    A successful transfer will emit a request to decrypt \lstinline{keyEncryptedSuccess}, a failed transfer will emit a request to decrypt \lstinline{keyEncryptedFailure}, which then results in decryption (if the keys were valid). The decryption of the keys will (usually) handled by an external oracle; see below for a proposal of the corresponding functionality.

    Cannot be called if \texttt{cancelAndDecrypt} was called before.

    \item \lstinline{cancelAndDecrypt}: Called by the receiver of the (payment) token whose address is implicit (\lstinline{to = msg.sender;}). 
    Verifies that the payer's (\lstinline{from}) address, receiver's (\lstinline{to}) address, and the encrypted keys match the corresponding call (with the same \lstinline{id}) to \lstinline{inceptTransfer}.
    Cancels the inceptTransfer call and emits a request to decrypt \lstinline{keyEncryptedFailure}.
    
    Cannot be called if \texttt{transferAndDecrypt} was called before.

    \item \lstinline{releaseKey}: Called by the decryption oracle with the decrypted key, to release (emit) the corresponding key.
\end{itemize}

% XXX TODO cancel for both?
% XXX TODO state preconditions on each phase

\pagebreak[5]
%%%%%%%%%%%%%%%%%%%%%%%%%%%%%%%%%%%%%%%%%%%%%%%%%%%%%%%%%%%%
\section{Workflow of a Secure Delivery-vs-Payment without External State}
\label{sec:dvp:workflow}
%%%%%%%%%%%%%%%%%%%%%%%%%%%%%%%%%%%%%%%%%%%%%%%%%%%%%%%%%%%

We describe the complete workflow of a secure Delivery-vs-Payment utilizing two smart contracts, implementing the \texttt{ILockingContract} and \texttt{IDecryptionContract}, respectively, and their interaction with a decryption oracle, see also Figure~\ref{fig:deliveryvspayment:sequence:full}. In this section we consider a single trusted decryption oracle. For the discussion of the decryption process and an the utilization of a distributed decryption process see Section~\ref{sec:dvp:decryptionoraclekeyformat}.

\subsection{Setup and Key Generation}

\begin{enumerate}
    \item The buyer generates the \lstinline{keyEncryptedSeller} ($E(S)$) and \lstinline{keyHashedSeller} ($H(S)$), using the \texttt{contract}-address of the desired \texttt{IDecryptionContract} and the \texttt{transaction} id as part of the key $S$

    \item The seller generates the \lstinline{keyEncryptedBuyer} ($E(B)$) and \lstinline{keyHashedBuyer} ($H(B)$), using the \texttt{contract}-address of the desired \texttt{IDecryptionContract} and the \texttt{transaction} id as part of the key $B$
\end{enumerate}

We suggest two alternative ways of key generation:
\begin{itemize}
    \item the buyer and seller may generate respective keys $K$ and use the decryption oracle's public encryption key to generate the pair $E(K)$, $H(K)$, while keeping $K$ secret, for $K=B$ or $K=S$, receptively, or
    
    \item the buyer and seller may call the decryption oracle's \texttt{requestEncryptedHashedKey} to generate the appropriate pairs $E(K)$, $H(K)$ without the requirement to (temporarily) store the key $K$. 
\end{itemize}
In any case, the respective other counterparty should use the decryption oracle's verify method to verify that $E(K)$ is associated with $H(K)$ and that both are associated with the correct contract and transaction.

As an implementation detail, it may be considered that key generation is handled directly by the smart contract - e.g., an event requesting the injection of appropriate encrypted/hashed keys and the decryption oracle reacting to this event.

\subsection{Buyer to \texttt{ILockingContract} (Open Transaction on Asset Chain)}

\begin{enumerate}[resume]
    \item The buyer executes on \texttt{ILockingContract} (asset) \lstinline{ILockingContract::inceptTransfer(uint256 id, int amount, address from, string keyHashedSeller, string keyEncryptedSeller)}. \label{itm:dvp:ilc:inceptTransfer}
\end{enumerate}

At this point $E(S)$ and $H(S)$ can be observed.

\subsection{Seller to \texttt{IDecryptionContract} (Open Transaction on Payment Chain)}

\begin{enumerate}[resume]
    \item Seller executes on \texttt{IDecryptionContract} (payment) \lstinline{IDecryptionContract::inceptTransfer(uint256 id, int amount, address from, string keyEncryptedBuyer, string encryptedKeySeller)}. \label{itm:dvp:idc:inceptTransfer}
\end{enumerate}

At this point $E(B)$ and $E(S)$ have been observed.

The buyer (receiver of the tokens on \texttt{ILockingContract}, payer of the tokens on \texttt{IDecryptionContract}) can now verify that the payment transfer has been incepted with the proper parameters. In particular, he can verify that \texttt{keyEncryptedBuyer} is associated with  \texttt{keyHashedBuyer}

\subsection{Verification against the Decryption Oracle}

If required, seller and buyer can verify the consistency of the encrypted/hashed keys.

\begin{enumerate}[resume]
    \item[$\bullet$] Buyer and/or seller call \texttt{verify} on the decryption oracle.
\end{enumerate}

\subsection{Buyer's cancellation option on \texttt{ILockingContract}}

\begin{enumerate}[resume] 
    \item[\ref{itm:dvp:ilc:confirmTransfer}*.] Buyer executes on \texttt{ILockingContract} (asset) \lstinline{ILockingContract::cancelTransfer(uint256 id, int amount, address from, string keyEncryptedSeller)}.
\end{enumerate}

This call can occur only after \lstinline{ILockingContract::inceptTransfer} (\ref{itm:dvp:ilc:inceptTransfer}) and before \lstinline{ILockingContract::confirmTransfer} (\ref{itm:dvp:ilc:confirmTransfer}) and will terminate this transaction.

\subsection{Seller to \texttt{ILockingContract} (Confirming on Asset Chain - Locking Asset)}

\begin{enumerate}[resume] 
    \item Seller executes on \texttt{ILockingContract} (asset) \lstinline{ILockingContract::confirmTransfer(uint256 id, int amount, address to, string keyHashedBuyer, string keyEncryptedBuyer)}. \label{itm:dvp:ilc:confirmTransfer}
\end{enumerate}

After this call, the asset will be locked by the \texttt{ILockingContract} for transfer to the buyer (upon successful payment) or transfer back to the seller (upon failed payment).

At this point $E(B)$ and $H(B)$ can be observed.

\subsection{Seller's cancellation option on \texttt{IDecryptionContract}}

\begin{enumerate}[resume]
    \item[\ref{itm:dvp:idc:transferAndDecrypt}*.] Seller executes on \texttt{IDecryptionContract} (payment)  \lstinline{IDecryptionContract::cancelAndDecrypt(uint256 id, address from, address to, string keyEncryptedBuyer, string encryptedKeySeller)}.
\end{enumerate}

This call can occur only after \lstinline{IDecryptionContract::inceptTransfer} (\ref{itm:dvp:idc:inceptTransfer}) and before \lstinline{IDecryptionContract::transferAndDecrypt} (\ref{itm:dvp:idc:transferAndDecrypt}). It will trigger the decryption of \texttt{encryptedKeySeller} (see \ref{itm:dvp:idc:TransferKeyRequested} below) and will terminate this transaction.

The seller can cancel the payment and obtain the key to re-claim the asset in case the buyer does not complete the payment.

\subsection{Buyer to \texttt{IDecryptionContract} (Completion on Payment Chain)}

\begin{enumerate}[resume]
    \item Buyer executes on \texttt{IDecryptionContract} (payment) \lstinline{IDecryptionContract::transferAndDecrypt(uint256 id, address from, address to, string keyEncryptedBuyer, string encryptedKeySeller)}. \label{itm:dvp:idc:transferAndDecrypt}
\end{enumerate}

\subsection{Completion of Transfer on \texttt{ILockingContract} (Completion on Asset Chain)}

\subsubsection{Upon Success:}

If the call to \lstinline{IDecryptionContract::transferAndDecrypt} resulted in a successful transfer of the (payment) tokens on the \texttt{IDecryptionContract}:
\begin{enumerate}[resume]
    \item The \texttt{IDecryptionContract} emits an event \texttt{TransferKeyRequested} with \lstinline{keyEncryptedBuyer} requesting decryption by the decryption oracle. \label{itm:dvp:idc:transferKeyRequested}

    \item The decryption oracle reacts to this event and decrypts \lstinline{keyEncryptedBuyer} to \lstinline{keyBuyer}, verifies that the event was issued by the corresponding \texttt{contract}, then calls \lstinline{IDecryptionContract::releaseKey} with \lstinline{keyBuyer}. \label{itm:dvp:idc:releaseKey}

    \item The buyer executes on \lstinline{ILockingContract::transferWithKey(uint256 id, string key)} with \lstinline{key = keyBuyer}. \label{itm:dvp:idc:transferWithKey}
\end{enumerate}

\subsubsection{Upon Failure:}

\begin{enumerate}[resume]
    \item[\ref{itm:dvp:idc:transferKeyRequested}*.] The \texttt{IDecryptionContract} emits an event \texttt{TransferKeyRequested} with \lstinline{keyEncryptedSeller} requesting decryption by the decryption oracle. \label{itm:dvp:idc:TransferKeyRequested}

    \item[\ref{itm:dvp:idc:releaseKey}*.] The decryption oracle reacts to this event and decrypts \lstinline{keyEncryptedSeller} to \lstinline{keySeller}, verifies that the event was issued by the corresponding \texttt{contract}, then calls \lstinline{IDecryptionContract::releaseKey} with \lstinline{keySeller}.

    \item[\ref{itm:dvp:idc:transferWithKey}*.] The seller executes on \lstinline{ILockingContract::transferWithKey(uint256 id, string key)} with \lstinline{key = keySeller}.
\end{enumerate}

\subsection{Communication between \texttt{IDecryptionContract} and Decryption Oracle}

The communication between the smart contract implementing \texttt{IDecryptionContract} and the decryption oracle is stateless.

The decryption oracle listens for the event \texttt{TransferKeyRequested}, which will show an \texttt{encryptedKey}, which is an encrypted document of the format suggested in the previous section.

The decryption oracle will decrypt the document \texttt{encryptedKey} without exposing the decrypted document \texttt{key}, verify that the event was issued from the contract specified in the decrypted \texttt{key}, and, in that case, submit the decrypted document \texttt{key} to the \texttt{releaseKey} function of \emph{that} contract.

\subsection{Complete Sequence Diagram}

The following sequence diagram summarizes the proposed delivery-versus-payment process; see Figure~\ref{fig:deliveryvspayment:sequence:full}.
\begin{figure}[hbtp]
    \centering
    \vspace{-4ex}
    \scalebox{0.50}{
  \begin{sequencediagram}
    \tikzstyle{inststyle}+=[bottom color=white, top color=white, rounded corners=3mm]
    \newthread[blue30]{s}{Seller of Asset}{}
    \newthread[red30]{b}{Buyer of Asset}{}
    \tikzstyle{inststyle}+=[bottom color=white, top color=white, rounded corners=0mm]
    \newinst[5]{a}{\shortstack{Asset Contract : \\ ILockingContract}}{}
    \newinst[2]{p}{\shortstack{Payment Contract : \\ IDecryptionContract}}{}
    \newinst[3]{o}{\shortstack{Decryption Oracle Operator : \\ DecryptionOracle}}{}

%    \postlevel
    \begin{sdblock}{Generation of \textbf{\color{green50}encrypted/hashed} keys.}{}
        \begin{call}{s}{\textbf{requestEncryptedHashedKey}(contract, id)}{o}{E(B), H(B)}
        \end{call}
        \begin{call}{b}{\textbf{requestEncryptedHashedKey}(contract, id)}{o}{E(S), H(S)}
        \end{call}
    \end{sdblock}
%    \postlevel

%    \begin{messself}{s}{\shortstack{\textbf{\color{green50}encrypt} B}}{E(B)}
%    \end{messself}
%
%    \prelevel
%
%    \begin{messself}{b}{\shortstack{\textbf{\color{green50}encrypt} S}}{E(S)}
%    \end{messself}

    \postlevel
    
    \begin{call}{b}{\textbf{inceptTransfer}(id, from, H(S), E(S))}{a}{\textit{TransferIncepted(id, from, to, H(S), E(S))}}
        \begin{messself}{a}{store H(S)}{}
        \end{messself}
    \end{call}

    \postlevel

    \begin{call}{s}{\textbf{inceptTransfer}(id, amount, from, E(B), E(S))}{p}{\textit{TransferIncepted(id, from, to, E(B), E(S))}}
    \end{call}

    \postlevel
    \begin{sdblock}{Verification of eligibility of encrypted keys}{Returns the address of the contract that is eligible to request decryption.}
        \begin{call}{s}{\textbf{verify}(E(S))}{o}{\texttt{IDecryptionContract} address, id, H(S)}
        \end{call}
    \end{sdblock}
    \postlevel
    
    \begin{call}{s}{\textbf{confirmTransfer}(id, to, H(B),E(B))}{a}{\textit{TransferConfirmed(id, from, to, H(B), E(B))}}
        \begin{messself}{a}{store H(B)}{}
        \end{messself}

        \postlevel

        \begin{messself}{a}{\textbf{\color{blue}lock asset tokens}}{}
        \end{messself}
    \end{call}

    \postlevel
    \begin{sdblock}{Verification of eligibility of encrypted keys}{Returns the address of the contract that is eligible to request decryption.}
        \begin{call}{b}{\textbf{verify}(E(B))}{o}{\texttt{IDecryptionContract} address, id, H(B)}
        \end{call}
    \end{sdblock}
    \postlevel

    \postlevel
    \begin{sdblock}{Alt: cancel (if payment is not initiated by buyer)}{\tiny K = S (sellers key)}
        \begin{call}{s}{\textbf{cancelAndDecrypt}(id, amount, to, E(B), E(S))}{p}{\textit{TransferKeyReleased(id, S)}}
        \begin{messcalldashed}{p}{\textit{TransferKeyRequested(id, E(S))}}{o}
            \postlevel
            \begin{messself}{o}{\shortstack[l]{\textbf{\color{red50}decrypt} E(S)\\ with private key}}{S}
            \end{messself}
            \postlevel
            \begin{messcall}{o}{\textbf{releaseKey}(id,S)}{p}
            \end{messcall}
            \prelevel
        \end{messcalldashed}
            \end{call}
    \end{sdblock}
    \begin{sdblock}{Alt: payment}{}
    \begin{call}{b}{\textbf{transferAndDecrypt}(id, amount, to, E(B), E(S))}{p}{\textit{TransferKeyReleased(id, K)}}
        \begin{call}{p}{\textbf{\color{blue}process payment order}}{p}{\shortstack[l]{E(K) = E(B) if successful,\\ E(K) = E(S) if failed}}
        \postlevel
        \end{call}
        \postlevel
        \begin{messcalldashed}{p}{\textit{TransferKeyRequested(id, E(K))}}{o}
            \postlevel
            \begin{messself}{o}{\shortstack[l]{\textbf{\color{red50}decrypt} E(K)\\ with private key}}{K}
            \end{messself}
            \postlevel
            \begin{messcall}{o}{\textbf{releaseKey}(id,K)}{p}
            \end{messcall}
            \prelevel
        \end{messcalldashed}
    \end{call}

    % Small hack. Emit key to both
    \prelevel
    \begin{messcalldashed}{p}{}{s}
    \end{messcalldashed}
    \prelevel
    \end{sdblock}

    \begin{sdblock}{Alt: success}{\tiny K = B (buyer key)}
        \begin{call}{b}{\textbf{transferWithKey}(id, B)}{a}{\textit{TokenClaimed(id)}}
            \begin{messself}{a}{\textbf{\color{green50}hash} K and \textbf{check} H(K) equals stored H(B)}{\textbf{\color{blue}transfer asset to buyer}}
            \end{messself}
        \end{call}
    \end{sdblock}
    \begin{sdblock}{Alt: failed}{\tiny K = S (sellers key)}
        \begin{call}{s}{\textbf{transferWithKey}(id, S)}{a}{\textit{TokenReclaimed(id)}}
            \begin{messself}{a}{\textbf{\color{green50}hash} K and \textbf{check} H(K) equals stored H(S)}{\textbf{\color{blue}transfer asset to seller}}
            \end{messself}
        \end{call}
    \end{sdblock}
  \end{sequencediagram}
    }
    \caption{Complete sequence diagram of a delivery-versus-payment transaction. % between a buyer and a seller (of the asset), interacting on the asset and the payment chain. Note the low involvement of the payment chain and payment operator.
    }
    \label{fig:deliveryvspayment:sequence:full}
\end{figure}

\newpage
%%%%%%%%%%%%%%%%%%%%%%%%%%%%%%%%%%%%%%%%%%%%%%%%%%%%%%%%%%%%
\section{Decryption Oracle and Key Format}
\label{sec:dvp:decryptionoraclekeyformat}
%%%%%%%%%%%%%%%%%%%%%%%%%%%%%%%%%%%%%%%%%%%%%%%%%%%%%%%%%%%%

The contracts rely on two keys, denoted by $B$ or $S$. Let $K$ denote any such key. The decryption of $K$ is done by an decryption oracle, which listens to messages that request decryption, and injects decrypted keys into the \texttt{releaseKey} method of the \texttt{IDecryptionContract}.

It is extremely important that the decryption oracle decrypts a key only if specific preconditions are met. The preconditions are
\begin{itemize}
    \item the request is issued from the eligible contract,

    \item the request is issued for the eligible transaction id (and the contract ensures that there cannot be two open transactions with the same id).
\end{itemize}

We propose a key format that allows to ensure that the decryption oracle releases the key $K$ only for the eligible contract / transaction.

While eligibility may imply state, we enforce it through a self-contained key format for the key $K$ and stateless verification protocol.

We propose the following elements:
\begin{itemize}
    \item A key document $K$ contains the \lstinline[language=Java]{contract} callback address of the contract implementing \texttt{IDecryptionContract}.

    \item A key document $K$ contains the \lstinline[language=Java]{transaction} specification (the id and possibly other information) of the transaction created by \texttt{IDecryptionContract::inceptTransfer}.

    \item The decryption oracle offers a stateless function \texttt{verify} that receives an encrypted key $E(K)$ and returns the \lstinline[language=Java]{contract} callback address (that will be used for \texttt{releaseKey} call), the \lstinline{transaction} detail that is stored inside $K$, and the hash $H(K)$ without returning $K$.

    The verify endpoint allows counterparties to confirm that a given encrypted key $E(K)$ maps to a specific contract and transaction, without revealing the plaintext $K$. This provides transparency without leaking key content.

    \item 
    \textbf{Security Property:} A decryption request $E(K)$ will only succeed if:
    \begin{enumerate}
        \item the embedded \lstinline[language=Java]{contract} and \lstinline[language=Java]{transaction} attributes in $K$ match the current call context, and

        \item and, if verified, passes $K$ to \texttt{releaseKey} of the callback \lstinline[language=Java]{contract} address found within the document $K$.
    \end{enumerate}
\end{itemize}

\pagebreak[5]
%%%%%%%%%%%%%%%%%%%%%%%%%%%%%%%%%%%%%%%%%%%%%%%%%%%%%%%%%%%%
\subsection{Key Format}

We propose the following XML schema for the document of the decrypted key:\footnote{The choice of XML is arbitrary and illustrative; other formats like JSON may be supported by an implementation.}
\begin{lstlisting}[language=XML]

<?xml version="1.0" encoding="utf-8"?>
<xs:schema attributeFormDefault="unqualified" elementFormDefault="qualified" targetNamespace="http://finnmath.net/erc/ILockingContract" xmlns:xs="http://www.w3.org/2001/XMLSchema">
    <xs:element name="releaseKey">
        <xs:complexType>
            <xs:simpleContent>
                <xs:extension base="xs:string">
                    <xs:attribute name="contract" type="xs:string" use="required" />
                    <xs:attribute name="transaction" type="xs:unsignedShort" use="required" />
                </xs:extension>
            </xs:simpleContent>
        </xs:complexType>
    </xs:element>
</xs:schema>
\end{lstlisting}
Here, the \texttt{contract}-attribute denotes a unique identification of a contract on a chain. This can be, for example, a CAIP-10 address, \cite{CIP7}. This attribute defines the contract for which the decryption should be performed. The \texttt{transaction}-attribute should contain the id of the transaction opened with \texttt{inceptTransfer} (where the \texttt{IDecryptionContract} ensures that two transaction ids cannot be open simultaneously).

A corresponding sample XML is shown below.
\begin{lstlisting}[language=XML]
<?xml version="1.0" encoding="UTF-8" standalone="yes"?>
<releaseKey contract="eip155:1:0x1234567890abcdef1234567890abcdef12345678" transaction="3141" xmlns="http://finnmath.net/erc/ILockingContract">
    <!-- random data -->
    zZsnePj9ZLPkelpSKUUcg93VGNOPC2oBwX1oCcVwa+U=
</releaseKey>
\end{lstlisting}
The decryption oracle should ensure that it performs decryption only for contracts matching the specification in the \texttt{contract}-attribute and transactions matching the specification in the \texttt{transaction}-attribute. This prevents replay attacks and the misuse of \texttt{inceptTransfer} and \texttt{cancelAndDecrypt} functions. The exact mechanism is an implementation detail of the decryption oracle.

%%%%%%%%%%%%%%%%%%%%%%%%%%%%%%%%%%%%%%%%%%%%%%%%%%%%%%%%%%%%‚
%\subsection{Decryption Oracle}
\subsection{Single Oracle Design: A Stateless API for the Decryption Oracle}

We consider the case of a single trusted decryption oracle and propose a stateless API for it.

By adding a method that provides encrypted keys, there is no requirement that the encryption method is known to anyone else, except the decryption oracle. A simple hashing method is sufficient. In addition, participants can verify the encrypted key without exposing the key by a dedicated \lstinline{verify} method.

The decryption oracle offers three stateless methods (endpoints):
\begin{itemize}
    \item \lstinline[language=Java]{requestEncryptedHashedKey(String contract, String transaction)}: internally generates $K$ (with the attributes provided), creates the encrypted key $E(K)$ and the hashed key $H(K)$ and returns the pair $E(K), H(K)$, without exposing $K$.

    \item \lstinline[language=Java]{verify(String encryptedKey)}: takes $E(K)$ and returns the corresponding \texttt{contract} and \texttt{transaction} fields (stored inside $K$) and $H(K)$, without exposing $K$.

    \item \lstinline[language=Java]{decrypt(String encryptedKey)}: takes $E(K)$, returns $K$, if the caller agrees with \lstinline[language=Java]{contract} found in $K$.
\end{itemize}

The decryption oracle owns a public/secret key pair for encryption/decryption of some key $K$.\footnote{Since $K$ will serve as a \textit{key} to the unlocking of tokens, we call $K$ key.} The key $K$ has the form
\begin{lstlisting}[language=Java]
    class ReleaseKey {
        @XmlAttribute(name = "contract")
        String contract;      // limit decryption request to eligible contract

        @XmlAttribute(name = "transaction")
        String transaction;   // limit decryption request to eligible transaction
        
        @XmlValue
        String value;        // secure random document
    }
\end{lstlisting}

Let $E(K)$ denote the encryption of $K$ and $H(K)$ denote a hash of $K$. We describe the detailed stateless functionality of the decryption oracle.

\subsubsection{Key Generation: \texttt{requestEncryptedHashedKey}}

\begin{lstlisting}[language=Java]
    EncryptedHashedKey requestEncryptedHashedKey(String contract, String transaction);
\end{lstlisting}
where
\begin{lstlisting}[language=Java]
    class EncryptedHashedKey {
        String encryptedKey;
        String hashedKey;
    }
\end{lstlisting}

The method \lstinline{requestEncryptedHashedKey} receives a contract id and a transaction specification. It then internally generates a random key $K$, incorporating the given \lstinline[language=Java]{contract} and \lstinline[language=Java]{transaction} attributes, performs encryption of $K$ to $E(K)$ and hashing of $K$ to $H(K)$, and returns $E(K)$ and $H(K)$ without exposing $K$.

\subsubsection{Key Verification: \texttt{verify}}

\begin{lstlisting}[language=Java]
    KeyVerification verify(String encryptedKey);
\end{lstlisting}
where
\begin{lstlisting}[language=Java]
    class KeyVerification {
        String contract;
        String transaction;
        String hashedKey;
    }
\end{lstlisting}

The method \lstinline{verify} internally decrypts the given $E(K)$ to $K$, extracts the \lstinline[language=Java]{contract} field and \lstinline[language=Java]{transaction} field from $K$, performs hashing of $K$ to $H(K)$, and returns the \lstinline[language=Java]{contract} field, the \lstinline[language=Java]{transaction} field and $H(K)$ without exposing $K$.

\subsubsection{Key Decryption: \texttt{decrypt}}

\begin{lstlisting}[language=Java]
    ReleaseKey decrypt(String encryptedKey);
\end{lstlisting}

The method \lstinline{decrypt} takes $E(K)$, internally decrypts it into $K$, verifies that the caller agrees with \lstinline[language=Java]{K.contract} and that the calling transaction agrees with \lstinline{K.transaction}. If verified, it returns $K$, otherwises it returns nothing / fails.

Here, \lstinline{encryptedKey} is an $E(K)$, the encryption of some $K$, e.g., as generated by \lstinline{requestEncryptedHashedKey}.

\subsection{Rationale for DvP}

For a secure DvP there will be two calls to \lstinline{requestEncryptedHashedKey} to obtain the encrypted / hashed success key (buyer's key $B$) and the encrypted / hashed failure key (seller's key $S$).

The decryption contract’s \lstinline{inceptTransfer} is initialized with the encrypted keys for success and failure of the payment.

The locking contract’s \lstinline{inceptTransfer}/\lstinline{confirmTransfer} is initialized with the hashed keys for success and failure of the payment.

If necessary, the seller and the buyer can verify that the contract keys are valid and consistent, i.e., that
\begin{itemize}
    \item $E(B)$ observed in \texttt{IDecryptionContract} has the hash $H(B)$ observed in \texttt{ILockingContract},

    \item $E(S)$ observed in \texttt{IDecryptionContract} has the hash $H(S)$ observed in \texttt{ILockingContract},

    \item \lstinline{B.contractId} and \lstinline{S.contractId} agrees with the contract id of the \texttt{IDecryptionContract}.
\end{itemize}
This can be achieved by the corresponding calls to the \texttt{verify} function of the decryption oracle.

\subsection{Distributed Oracle Design: Threshold Decryption}

The decryption oracle does not need to be a single trusted entity. Instead, a threshold decryption scheme \cite{DesmedtFrankel1990} can be employed, where multiple oracles perform partial decryption, requiring a quorum of them to reconstruct the secret key. This enhances security by mitigating the risk associated with a single point of failure or trust.

In such cases, each participating decryption oracle will observe the decryption request from an emitted \texttt{TransferKeyRequested} event, and subsequently call the \texttt{releaseKey}(id,$K_{i}$) method with a partial decryption result $K_{i}$, see Figure~\ref{fig:dvp:decryptionoracle:distributed}.

Once sufficient partial decryptions $K_{i}$ have been submitted to the decryption contract, the key $K$ can be reconstructed. The reconstruction of $K$ is an implementation detail. An option is that the decryption contract performs the reconstructions and emits a final \texttt{TransferKeyReleased} event with the reconstructed key. If on-chain reconstruction is considered computationally too intense, another option is that the partial decryptions are merely published and the reconstruction is left to the user.

To allow both implementation variants, the \texttt{TransferKeyReleased}-event has a \texttt{sender} argument that exhibits the caller of the \texttt{releaseKey}-function. This allows to distinguish partial decryptions from different external oracles, while a final reconstruction could show the decryptions contracts address as the sender in the \texttt{TransferKeyReleased}-event.

\begin{figure}[hbtp]
    \centering
    \scalebox{0.50}{
  \begin{sequencediagram}
    \tikzstyle{inststyle}+=[bottom color=white, top color=white, rounded corners=3mm]
    \newthread[blue30]{u}{User}{}
    \tikzstyle{inststyle}+=[bottom color=white, top color=white, rounded corners=0mm]
    \newinst[5]{c}{\shortstack{Payment Contract : \\ IDecryptionContract}}{}
    \newinst[5]{o}{\shortstack{Decryption Oracle\\0x1234}}{}
    \newinst[2]{o2}{\shortstack{Decryption Oracle\\0x5678}}{0x5678}
    \newinst[2]{o3}{\shortstack{Decryption Oracle\\0xabcd}}{}

    \begin{sdblock}{variant with shared/distributed decryption oracles}

    \begin{call}{u}{\textbf{transferAndDecrypt(\ldots)}}{c}{\textit{}}

    \begin{call}{c}{preconditionsCheck()}{c}{status}
    \end{call}
    
    \begin{sdblock}{Alt: if status valid}{}

        \begin{messcalldashed}{c}{\textit{TransferKeyRequested(id, \textbf{\color{blue}encryptedKey)}}}{o}
            \postlevel
            \begin{messself}{o}{\shortstack[l]{\textbf{\color{red50}decrypt}    \textbf{\color{blue}encryptedKey}\\ with private key}}{\textbf{\color{green50}key$_\mathbf{1}$}}
            \end{messself}
            \begin{messcall}{o}{\textbf{releaseKey}(0x1234,id,\textbf{\color{green50}key$_\mathbf{1}$})}{c}
            \end{messcall}
            \prelevel
        \end{messcalldashed}

        \prelevel
        \prelevel
        \prelevel
        \prelevel
        \prelevel
        
        \begin{messcalldashed}{c}{}{o2}
            \postlevel
            \begin{messself}{o2}{\shortstack[l]{\textbf{\color{red50}decrypt}    \textbf{\color{blue}encryptedKey}\\ with private key}}{\textbf{\color{green50}key$_\mathbf{2}$}}
            \end{messself}
            \postlevel
            \postlevel
            \begin{messcall}{o2}{\textbf{releaseKey}(0x5678,id,\textbf{\color{green50}key$_\mathbf{2}$})}{c}
            \end{messcall}
        \end{messcalldashed}

        \prelevel
        \prelevel
        \prelevel
        \prelevel
        \prelevel
        \prelevel
        \prelevel
        \prelevel
        
        \begin{messcalldashed}{c}{}{o3}
            \postlevel
            \begin{messself}{o3}{\shortstack[l]{\textbf{\color{red50}decrypt}    \textbf{\color{blue}encryptedKey}\\ with private key}}{\textbf{\color{green50}key$_\mathbf{3}$}}
            \end{messself}
            \postlevel
            \postlevel
            \postlevel
            \postlevel
            \begin{messcall}{o3}{\textbf{releaseKey}(0xabcd,id,\textbf{\color{green50}key$_\mathbf{3}$})}{c}
            \end{messcall}
        \end{messcalldashed}

        \begin{messself}{c}{key reconstruction}{\textbf{\color{green50}key}}
        \end{messself}

    \begin{messcalldashed}{c}{TransferKeyReleased(self, id, key)}{u}
    \end{messcalldashed}

    \prelevel
    \end{sdblock}
    
    \begin{sdblock}{Alt: else}{}

        \prelevel
        \begin{messself}{c}{}{}
        \end{messself}
        
        % Small hack. Emit key to both
        \begin{messcalldashed}{c}{Failure}{u}
        \end{messcalldashed}
    
    \prelevel
    \prelevel
    
    \end{sdblock}

    \prelevel

    \end{call}

    \end{sdblock}

  \end{sequencediagram}
    }
    \caption{Shared/distributed decryption with a threshold decryption scheme.
    }
    \label{fig:dvp:decryptionoracle:distributed}
\end{figure}

\clearpage
\section{Remarks}
\label{sec:dvp:remarks}

\subsection{Encryption versus Hashing}

The above scheme requires the use of an encrypted key $E(K)$ and a corresponding hashed key $H(K)$. This requires that the participants can check the consistency of the pair $E(K), H(K)$.

As encryption can be performed with a public key, in theory the hashing could be replaced by encryption. In that case the protocol simplifies slightly with $H(K) = E(K)$. The use of a separate hashing method is for practical reasons only, as on-chain encryption may be expensive.

However, the participants still need to check that $E(K)$ represents a proper encryption of an eligible key, i.e., the \texttt{verify} step will still be necessary to verify that $K$ is a propper key for the specific contract and transaction.

\subsection{Key Generation}

The protocol suggests that the generation of the buyer's key $B$ is performed/requested by the seller, and that the generation of the seller's key $S$ is performed/requested by the buyer.

This is intentional to avoid a \textit{replay-attack} where it would be possible to reuse a previously observed key. It is in the interest of the seller that the hash of the buyer's key is not that of a previously observed key, and vice versa.

The \texttt{transaction}-attribute of the key format has to be used to eliminate the risk of a replay-attack.

The decryption oracle may offer a convenient service to generate appropriate encrypted keys $E(K)$ and hashed $H(K)$. This represents an important improvement in overall security, as it removes the need that $K$ exists outside the decryption oracle prior to the decryption step.

\subsection{Security Considerations}

\subsubsection{Preventing Replay Attacks}

To prevent replay attacks that allow to perform a decryption of some $E(K)$ through a different contract or a different transaction the contract address and the transaction id should be part of the key $K$ \emph{and} the description oracle should perform decryption only for eligible keys.

For example, without such a check, a simple attack is to initialize a payment transaction with \lstinline{inceptTransfer(id, from, E(Y), E(X))} followed by a call to \lstinline{cancelTransfer(id, from, E(Y), E(X))}. This will result in a decryption of the presumed failure-key $X$. If $E(X)$ is the success-key of some other transaction, this then allows the attacker to obtain the asset without payment. If $E(X)$ is the failure-key of some other transaction, this then allows to re-claim the asset after a buyer has paid.

This attack cannot occur if the locking of the asset with \lstinline{ILockingContract::confirmTransfer} is performed after the call to \lstinline{IDecryptionContract::inceptTransfer} and the \lstinline{IDecryptionContract} does not allow to open a transaction with the same id.

\subsubsection{Consensus Decryption Oracle via Threshold Encryption}

The keys that are decrypted by the decryption oracle are useless to a third person, as smart contracts can still ensure that a buyer or seller, but no other party, can receive the asset(s). However, the decryption oracle has to be a trusted entity. It should be noted that it is not necessarily a central trusted entity. First, an appropriate decryption oracle could be chosen out of many on a per-contract basis.
In addition, an option could be to implement a $k$-out-of-$n$ threshold decryption. Such an algorithm allows to encrypt a message using $n$ public keys, where the decryption requires the decryption using at least $k$ secret keys. 

\bigskip

Table~\ref{tbl:dvp:security_defenses} summarizes threat vectors and the protocol’s corresponding defense mechanisms.

\bigskip

\begin{table}[hbtp]
    \centering
    \footnotesize
    \renewcommand{\arraystretch}{1.4}
    \begin{tabular}{|p{0.40\textwidth}|p{0.55\textwidth}|}
        \hline
        \textbf{Threat} & \textbf{Defense Mechanism} \\
        \hline
        Replay of key usage in other contracts or chains & Embedded \texttt{contract} and \texttt{transaction} id in encrypted key $K$; enforced by oracle before release \\
        \hline
        Key reuse due to deterministic generation & Randomized key payload and transaction-scoped identifiers prevent key collision or duplication \\
        \hline
        Malicious or compromised decryption oracle & Optional threshold decryption across multiple oracles \\
        \hline
        Observable keys leaked in mempool or logs & Encrypted keys $E(K)$ are never revealed before proper contract call and validation \\
        \hline
        Unauthorized contract triggering \texttt{releaseKey()} & Oracle enforces that only the contract embedded in $K$ may request decryption \\
        \hline
        Cancel call races or replays after failure & Cancel path is handled explicitly via embedded cancel key $S$, scoped to transaction context \\
        \hline
    \end{tabular}
    \caption{Security threats and corresponding defenses in the DvP protocol}
    \label{tbl:dvp:security_defenses}
\end{table}

\subsection{Pre-Trade versus Post-Trade}

The distinction between pre-trade and post-trade failure is central to financial settlement protocols, where it affects whether a failed exchange is interpreted as a normal cancellation or a counterparty default. This distinction is formally addressed in traditional Delivery-versus-Payment models, such as the BIS report on securities settlement \cite{bis1992dvp}.

In our protocol, it might be considered a \textit{failure to pay} if the buyer does not initiate \lstinline{IDecryptionContract::transferAndDecrypt}. However, this depends on interpreting which actions are deemed part of the trade inception phase and which are regarded post-trade.

Assume that we interpret the first three transactions, i.e.,
\begin{enumerate}
    \item \lstinline{ILockingContract:inceptTransfer},
    \item \lstinline{IDecryptionContract:inceptTransfer},
    \item \lstinline{ILockingContract:confirmTransfer},
\end{enumerate}
as being pre-trade, manifesting a \textit{quote} and the seller's intention to offer the asset for the agreed price.

We may interpret the fourth transaction, i.e.,
\begin{enumerate}[resume]
    \item \lstinline{IDecryptionContract::transferAnDecrypt},
\end{enumerate}
as manifesting the trade inception and initiating the post-trade phase. So we could interpret this transaction as a \lstinline{IDecryptionContract:confirmTransfer} that also immediately triggers the completion of the transaction.

If we take the two interpretations above, then \lstinline{IDecryptionContract::transferAnDecrypt} marks the boundary of trade event and post-trade transactions. This interpretation implies that a lack of \lstinline{IDecryptionContract::transferAnDecrypt} does not represent a \textit{failure to pay} and a \lstinline{IDecryptionContract::cancelAndDecrypt}, initiated by the seller, has the straightforward interpretation of invalidating a pre-trade quote (or offer).

It may be disputed if a failed payment resulting from a \lstinline{IDecryptionContract::transferAnDecrypt} represents a failed inception of the trade (pre-trade) or a failure to pay (post-trade). In any case, the seller is not facing the risk of losing the asset without a payment, and the buyer is not facing the risk of losing the payment without a delivery of the asset.

Note that introducing a locking scheme for the payment is either unnecessary (because the transfer can be completed immediately) or raises a corresponding possibility of a failure-to-deliver (if the asset locking occurs after the payment locking).

\subsection{Embedding an Existing HTLC (optional wrapper)}

An existing HTLC contract can be embedded into the key-based DvP with and \lstinline{IDecryptionContract}. The use of an \lstinline{IDecryptionContract} with an existing HTLC creates benefits, as it allows invalidating the receiver’s option induced by the HTLC timeout and mitigating the HTLC’s compromising race conditions. It adds the benefit of the cancellation feature, with a mild drawback: liquidity remains locked until the HTLC refunds at timeout.

We assume that on one of the two chains an HTLC implementation exists that supports:
\begin{enumerate}
    \item \textbf{lock} the token with a timeout,
    \item  \textbf{complete} the transfer by presenting a preimage (secret) matching a given hash,
    \item  \textbf{refund} if the preimage was not presented after the timeout.
\end{enumerate}

\subsubsection{Implementation}

\paragraph{ILockingContract:} The ILockingContract wraps the HTLC as follows:
\begin{itemize}
    \item \lstinline{inceptTransfer} triggers the \textbf{lock} on the HTLC.
    
    \item \lstinline{transferWithKey} with a successKey (B) \textbf{complete}s the HTLC’s transfer (just passes the successKey (B) as the HTLC preimage).
    
    \item \lstinline{transferWithKey} with a failureKey (S) finalizes cancellation in the wrapper.
    The underlying HTLC then \textbf{refund}s at timeout $T_{2}$; this invalidates the receiver’s option.
\end{itemize}
This implementation is optional, but may provide additional benefits; see below.

\paragraph{IDecryptionContract:} The implementation of \lstinline{IDecryptionContract} has to be initialized with an additional timeout parameter $T_{1}$ being sufficiently smaller than the HTLC timeout $T_{2}$. The methods are then implemented as follows:
\begin{itemize}
    \item the \lstinline{cancelAndDecrypt} method decrypts the \lstinline{failureKey} $S$ and blocks further operations (i.e., disallowing decryption of the \lstinline{successKey} $B$).
    
    \item the \lstinline{transferAndDecrypt} method
    decrypts \lstinline{failureKey} $S$ if called after $T_{1}$, otherwise it performs the transfer and decrypts \lstinline{successKey} $B$ upon successful transfer and the \lstinline{failureKey} $S$ upon a reverted transfer.
\end{itemize}

As in Section~\ref{sec:dvp:IDecryptionContract} and Figure~\ref{fig:deliveryvspayment:sequence:full}, \lstinline{transferAndDecrypt} and \lstinline{cancelAndDecrypt} are mutually exclusive (first one wins).

As for a classical HTLC, its secret (the \lstinline{successKey} $B$ for the \lstinline{ILockingContract}) is only visible if the transaction on the \lstinline{IDecryptionContract} was successful. 

The novelty is that the party who locked on the HTLC
obtains a cancellation option that invalidates the receiver’s option and thereby ensures that the HTLC timeout elapses.

\subsubsection{Implementation Variants}

Once the \lstinline{failureKey} $S$ is released, the \lstinline{successKey} $B$ will not be released; therefore using the \lstinline{ILockingContract} as a strict wrapper is optional. Once the \lstinline{failureKey} $S$ is released, the HTLC will inevitably reach its timeout; the \lstinline{failureKey} $S$ will not be used on the HTLC in this case.

Implementing an \lstinline{ILockingContract} allows to improve onchain state transparency, as we can put the contract is a secure cancellation state, and is advisable if one plans to transition to a key‑based cancellation implementation (without timeout) later.

An obviously less secure alternative for the \lstinline{IDecryptionContract} is to keep the standard implementation and rely on an external time-oracle to call \textit{cancelAndDecrypt} upon timeout $T_{1}$.

\subsubsection{Benefits}

Combining a classical HTLC with \lstinline{IDecryptionContract} does not provide the full benefits of key-based DvP; however, it still brings improvements:
\begin{itemize}
    \item The locked token can be canceled at any time, invalidating the receiver’s option embedded in the HTLC timeout (the token returns only after the original HTLC timeout expires, but the option is invalidated).

    \item Due to the possibility of invalidating the option, the timeout may be chosen larger, making the contract less prone to race conditions. A longer timeout no longer increases the value of the receiver's option but still imposes liquidity costs.
\end{itemize}

\subsection{Payment Locking versus Asset Locking}

The realization of delivery-versus-payment with \lstinline{IDecryptionContract} and \lstinline{ILockingContract} requires locking on one of the two chains only. The locked token will be reserved for the duration of the transaction, while the transfer on the other chain succeeds or fails instantaneously.

Depending on the application, it may be favorable to have the locking contract on a specific side, e.g., to have asset locking instead of cash locking to avoid cash being tied up for a long time.

\clearpage
\section{Multi-Party Delivery versus Payment}
\label{sec:dvp:multiparty}

\subsection{Locking is a Feature}

In Delivery-versus-Payment (DvP) protocols at least one token must be locked to ensure atomicity, even if only for a short period during the transaction.

While locking may appear as an inconvenient necessity, it is in fact a feature that becomes valuable in the construction of conditional trades or multi-party DvPs.

A multi-party delivery versus payment is a valuable trade feature. Consider, for example, the case where counterparty A wishes to buy a token $Y$ (e.g., a bond) from counterparty C, but in order to fund this transaction, counterparty A wishes to sell a token $X$ (e.g., another bond) to counterparty B. However, A does not want to sell bond $X$ if the purchase of $Y$ fails. A multi-party DvP allows these two transactions to be bound into a single atomic unit.
The liquidity required for the combined transaction is then only the net of the two individual transactions.

\medskip

While for a two-party DvP with two tokens only one token requires locking—and hence a DvP can be constructed without locking on the cash chain—a three-party DvP with three tokens in general requires the ability to lock all three tokens.

If $n$ parties wish to perform bilateral transactions atomically, there are \textit{at least} $m := 2 \cdot (n - 1)$ transactions, of which $m-1$ require locking. The last one can operate directly, and its success or failure decides whether the other locks are released or reverted.

This highlights that locking is not just a constraint, but a required feature to enable advanced and economically meaningful protocols.

\subsection{Mutli-Party DvP with \lstinline{IDecryptionContract}}

A multi-party DvP can be created elegantly by combining multiple two-party DvPs, for example based on the ERC-7573 protocol.

The procedure is simple: instead of finalizing the respective two-party DvP by a call to \lstinline{transferAndDecrypt}, the two-party DvP is first confirmed with a call to \lstinline{confirm}, leaving the finalization open.

At any time, any party can call \lstinline{cancelAndDecrypt} to release the failure key and revert all lockings.

Once all parties are linked with their respective two-party DvPs, a single call to \lstinline{transferAndDecrypt} performs locking of the token implementing the \lstinline{IDecryptionContract} and releases either the success key on success or the failure key on failure.

\subsubsection{Initiation and Finalization}

The counterparty that initiates the multi-party DvP by making the first call
to the \lstinline{IDecryptionContract} is the one that is allowed to finalize it via \lstinline{transferAndDecrypt}, the other may cancel via \lstinline{cancelAndDecrypt}.

\subsubsection{Sequence Diagram}

\begin{figure}[hbtp]
    \centering
    \scalebox{0.40}{
        \includegraphics{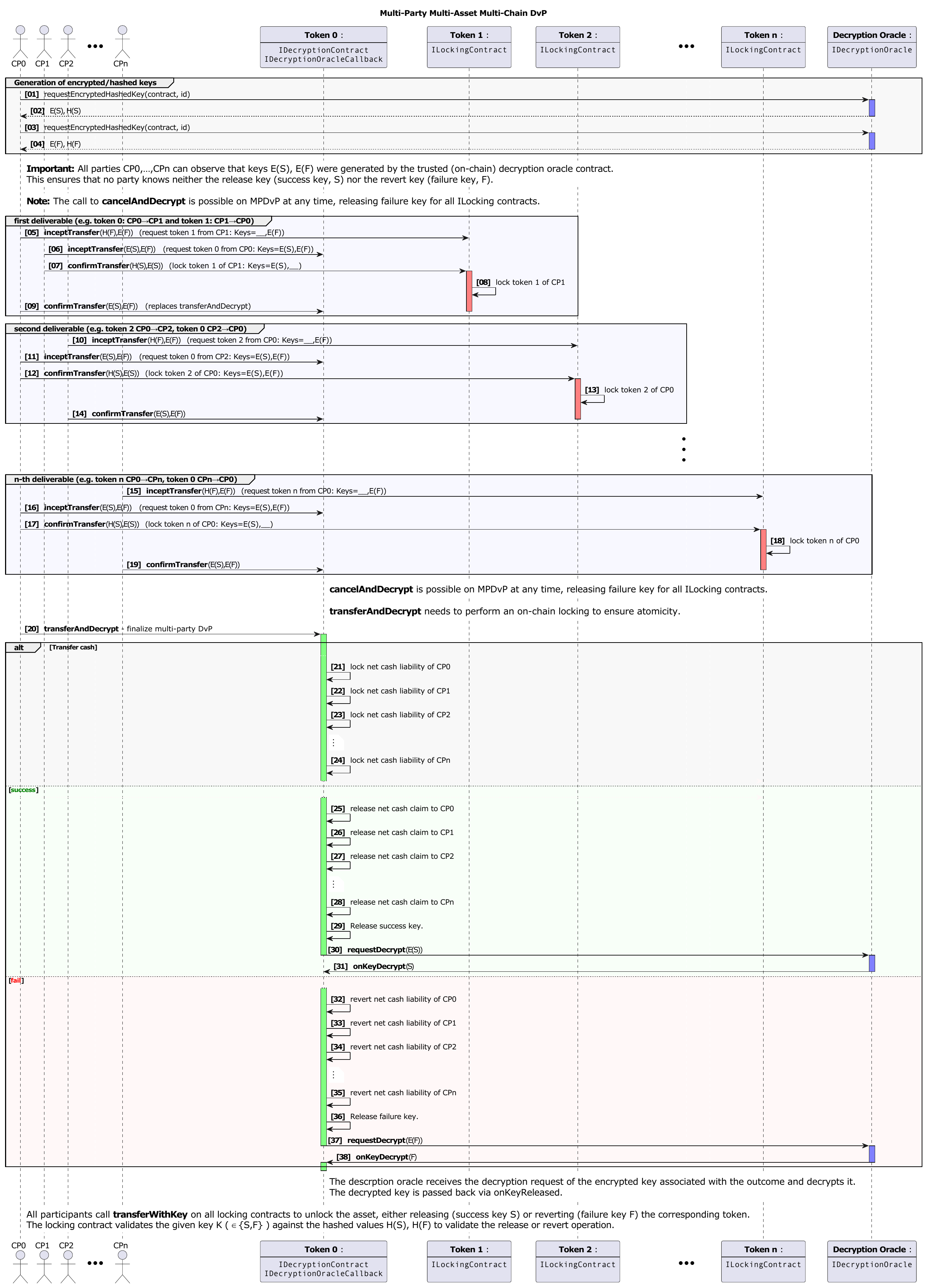}
    }
    \caption{Multi-Party Delivery versus Payment.
    }
    \label{fig:dvp:multi-party-dvp}
\end{figure}

In Figure~\ref{fig:dvp:multi-party-dvp} we depict the corresponding sequence diagram of a multi-party DvP via ERC-7573.
Note that the individual DvP may come in two different flavors depending on which counterparty is the receiver of the token on the \lstinline{IDecryptionContract}.

The diagram depicts a multi-party dvp with n+1 counterparties trading n+1 tokens out of which
the DvPs are bound by the contract on token 0.

Note: The more general case of N counterparties trading
M tokens is just a special case where we enumerate all combination as new counterparties and new tokens.

\clearpage
\section{Conclusion}
\label{sec:dvp:conclusion}

We proposed a decentralized transaction scheme that allows one to realize a secure delivery-versus-payment across two (blockchain) infrastructures without the requirement to hold state outside the chains. The payment system requires minimal integration and remains decoupled from the asset transaction. We proposed a key format that prevents replay attacks.

The proposed decryption oracle can also be used to resolve race-conditions in other applications.

Compared to other delivery-versus-payment schemes, the main advantage of the present approach is:
\begin{itemize}
    \item \textbf{No Intermediary Service Holding State}: Hashes or keys are not required to be stored by a third-party service. Hence, there is no additional point of failure. The decryption oracle operator's public key serves as the encryption key.
\end{itemize}

Additional advantages, some shared with HTLC-based schemes, include:
\begin{itemize}
    \item \textbf{No Centralized Key Generation}: Keys can be generated mutually by the trading parties at the trade inception phase and will not be needed afterwards. Centralized key generation is a convenient option.
    \item \textbf{No Timeout Scheme}: The transaction is not required to complete in a given time window, hence no timeout. The timing is up to the two counterparties. Either the buyer initiates the payment (via \lstinline{IDecryptionContract::transferAndDecrypt}) or, in case of absence of payment initiation, the seller cancels the offer and re-claims the asset (via \lstinline{IDecryptionContract::cancelAndDecrypt}). The absence of time-outs remove the possibility of unwanted race-conditions.
    \item \textbf{No Coupling}: The payment chain and the payment chain operator do not need any knowledge of the associated asset. They only offer the possibility to observe the transaction state, which then triggers the decryption.
    \item \textbf{Lean Interaction}: The function workflow is structured and only consists of three main interactions:
    \begin{enumerate}
        \item generate encrypted keys and lock assets,

        \item send payment order with encrypted keys,
        
        \item retrieve decrypted keys and unlock assets.
    \end{enumerate}
\end{itemize}

Our scheme is particularly suited for programmable money environments, tokenized assets, and settlement systems involving central bank digital currencies (CBDCs), where timing constraints and external coordination are infeasible or undesirable.

The proposed smart contract interfaces are available as ERC 7573, \cite{erc7573}. An open source reference implementation of the decryption oracle is available at \cite{finmath-decyption-oracle}

Future work includes deployment in real DvP transactions and formal analysis of liveness under network faults.

\clearpage
%%%%%%%%%%%%%%%%%%%%%%%%%%%%%%%%%%%%%%%%%%%%%%%%%%%%%%%%%%%%%%%%%%%%%
%\section{References}
%%%%%%%%%%%%%%%%%%%%%%%%%%%%%%%%%%%%%%%%%%%%%%%%%%%%%%%%%%%%%%%%%%%%%

\bibliographystyle{unsrt} % abbrvnat or unsrt
\bibliography{localbibliography.bib}

\clearpage
\appendix

\section{Double-Locking on two Contracts}
\label{sec:dvp:doublelocking}

In the following $S$, $B$, $C$, $X$ and $Y$ are secret documents (e.g., long random sequences) and $H(S)$, $H(B)$, $H(C)$, $H(X)$, $H(Y)$ are \textit{hashes} of the respective document. For a given $K$ the calculation of $H(K)$ is easy, for a given $H(K)$ the determination of $K$ is hard (presumed impossible)

We consider two contracts, \textit{asset} and \textit{payment}, each running on a separate chain. The names are purely illustrative. The aim is to perform a transfer on \textit{asset} if and only if a transfer on \textit{payment} was successful.

There shall be parties, the \textit{buyer} and the \textit{seller}. The buyer is receiving the asset and delivering the payment.
The seller is delivering the asset and receiving the payment.

The contracts can store values, calculate hashes, and compare them with previously stored values. Execution of transfer is conditional to a successful comparison.
The contracts can verify the caller of a function.

\subsection{Steps to Create the Double Locking}

\subsubsection*{Key Generation}

\begin{enumerate}
    \setcounter{enumi}{-1}
    \item \textbf{Key Generation}: Buyer knows $S$, $Y$. Seller knows $B$, $C$, $X$.
\end{enumerate}

\subsubsection*{Partially Locking the Asset}

\begin{enumerate}[resume]
    \item \textbf{Buyer}: \texttt{asset.incept(H(S), H(Y))} 
    
    Buyer knows $S$ and $Y$. The asset contract internally stores $H(S)$ and $H(Y)$.

    \item \textbf{Seller}: \texttt{asset.confirm(H(B), H(C), H(X))} \label{itm:dvp:dbllck:assetConfirm}
    
    Seller knows $B$ and $C$ and $X$. The asset contract internally stores $H(B)$ and $H(C)$ and $H(X)$.

    The asset contract moves the asset from the seller account to a lock.
\end{enumerate}

From this point on and until 5), the seller can call the cancel function on asset at any time.
\begin{enumerate}[resume]
    \item[\ref{itm:dvp:dbllck:assetConfirm}*] \textbf{Seller}: \texttt{asset.cancel(C)}

    Transfers the asset back to the seller. The buyer can observe C (i.e., the buyer knows C, once the seller cancels).
\end{enumerate}

The asset is now locked with a cancel option:
\begin{itemize}
    \item Seller can cancel by presenting C (asset is transferred back to seller).

    \item Buyer has no stake in the transaction yet.
    
    \item Buyer can retrieve asset with B (but does not know B yet).

    \item Buyer can remove Seller’s cancellation option by presenting (X,Y) (but does not know X yet).
\end{itemize}

\subsubsection*{Locking the Payment}

\begin{enumerate}[resume]

% 3
    \item \textbf{Buyer}: \texttt{payment.incept(H(B), H(S), H(C), H(X))} \label{itm:dvp:dbllck:paymentIncept}
    
    The payment contract internally stores H(B) and H(S) and H(C) and H(X).

% 3'
    \item[\ref{itm:dvp:dbllck:paymentIncept}*] \textbf{Buyer}: \texttt{payment.cancel(H(B), H(S), H(C), H(X)}
    
    The buyer can cancel the previous incept if the seller does not confirm. In this case the seller will use C to cancel the asset locking.

% 4

    \item \textbf{Seller}: \texttt{payment.confirm(H(B), H(S), H(C), X)} \label{itm:dvp:dbllck:paymentConfirm}

    The payment contract verifies that $H(X)$ equals the previously stored $H(X)$.
    
    The payment is moved from the buyers account to the lock. The buyer can no longer cancel the payment locking without presenting $C$.
\end{enumerate}

The payment is locked:
\begin{itemize}
    \item Seller can retrieve the payment with (B,Y), but Y has not been observed yet.

    \item Buyer can cancel the payment with C (if Seller cancels asset with C).
    
    \item The value of X has been observed.
\end{itemize}

Buyer has stake in the transaction, Seller can cancel with C. Buyer cannot cancel (without Seller’s cancellation), hence he likes to remove Seller’s cancellation option. He is next. Now, that he has observed X he will remove the Seller’s cancellation option asap. He could not do this before, because this requires X.
    
\subsubsection*{Locking the Asset}

Continuation on asset contract requires that $X$ has been observed in the previous step on the payment contract.

\begin{enumerate}[resume]
    \item \textbf{Buyer}: \texttt{asset.lock(H(B),H(S), X, Y)} \label{itm:dvp:dbllck:assetLock}
\end{enumerate}

The asset is locked.

\begin{itemize}
    \item Buyer can retrieve asset with B (but does not know B yet).
    
    \item Seller can retrieve asset with S (but does not know S yet).

    \item Seller cannot cancel with C anymore.
    
    \item Y has been observed. Seller can retrieve payment.
\end{itemize}

Both parties have stake in the transaction. The symmetric double-locking is established.

\subsubsection*{Completing the Transaction: Payment}

Continuation on Payment Contract requires that Y has been observed in the previous step on the asset contract

Either of the following two will finalize the transaction.
\begin{enumerate}[resume]
    \item \textbf{Seller}: \texttt{payment.retrieve(B,Y)} \label{itm:dvp:dbllck:sellerPaymentRetrieve}
    
    Seller gets payment, exposes B, Buyer can take asset.

    \item[\ref{itm:dvp:dbllck:sellerPaymentRetrieve}*] \textbf{Buyer}: \texttt{payment.retrieve(S)} \label{itm:dvp:dbllck:buyerPaymentRetrieve}
    
    Buyer gets payment, exposes S, Seller can take asset.
\end{enumerate}

\subsubsection*{Completing the Transaction: Asset}

Continuation on Asset Contract requires that either $B$ or $S$ has been observed in the previous step on the asset contract.

\begin{enumerate}[resume]
    \item \textbf{Buyer}: \texttt{asset.retrieve(B)} \label{itm:dvp:dbllck:buyerAssetRetrieve}
    
    Buyer can take asset, knows B.

    \item[\ref{itm:dvp:dbllck:buyerAssetRetrieve}*] \textbf{Seller}: \texttt{asset.retrieve(S)} \label{itm:dvp:dbllck:sellerAssetRetrieve}
    
    Seller gets asset, knows S.
\end{enumerate}

\subsection{Race Conditions}

The above protocol exhibits three race conditions:
\begin{enumerate}
    \item The cancellation of the payment by the buyer in \ref{itm:dvp:dbllck:paymentIncept}* could conflict with the confirmation of the payment by the seller in \ref{itm:dvp:dbllck:paymentConfirm}. If the first call goes through the buyer receives the payment back. If the second call is observed, the buyer observes $X$ and can lock the asset with \ref{itm:dvp:dbllck:assetLock}.

    \item The cancellation of the asset by the seller in \ref{itm:dvp:dbllck:assetConfirm}* could conflict with the locking of the asset by the buyer in \ref{itm:dvp:dbllck:assetLock}. If the first call goes through the seller receives the asset back. If the second call is observed, the seller observes $Y$ and can retrieve the payment with \ref{itm:dvp:dbllck:sellerPaymentRetrieve}.

    \item The retrieval of the payment by the seller in \ref{itm:dvp:dbllck:sellerPaymentRetrieve} could conflict with the retrieval of the payment by the buyer in \ref{itm:dvp:dbllck:buyerPaymentRetrieve}*. If both calls are observed, we observe $B$ and $S$ simultaneously, leaving it open who gets the asset.
\end{enumerate}

While hashes allow to synchronize the executions across two chains, we are left with the issue that simultaneous function calls on the same chain could allow the observation of their arguments, which may compromise the scheme.

%
% Keywords
%
\section*{Notes}

\subsection*{Keywords}

Delivery vs Payment,
DvP,
Settlement,
Atomic Swap,
Hashed Timelock Contract,
HTLC,
Smart Contract,
Blockchain,
ERC 7573

\subsection*{JEL Codes}

E42, G20

\subsection*{Suggested Citation}

\textsf{Fries, Christian P. and Kohl-Landgraf, Peter:
\textit{\articletitle}}
(November 9, 2023). Available at SSRN: \url{https://ssrn.com/abstract=4628811} or \url{http://dx.doi.org/10.2139/ssrn.4628811}

\bigskip\bigskip

\noindent
For updates see \url{https://ssrn.com/abstract=4628811}.
\end{document}